\documentclass{aa}
\usepackage[english]{babel}
\usepackage{graphicx}
\usepackage{txfonts}

\newcommand{\sax}{{\it BeppoSAX}}

\newcommand{\swf}{{\it Swift}}

\newcommand{\egret}{{EGRET}}
\newcommand{\agile}{{\it AGILE}}
\newcommand{\glast}{{\it Fermi}}
\newcommand{\mrkb}{Mrk~501}
\newcommand{\mrka}{Mrk~421}
\begin{document}
\title{SSC radiation in BL Lac sources,\\the end of the tether}
  \subtitle{}

\author{A.~Paggi\inst{1}, F.~Massaro\inst{2}, V.~Vittorini\inst{3}, A.~Cavaliere\inst{1}, F.~D'Ammando\inst{1,3}, F.~Vagnetti\inst{1}, M.~Tavani\inst{1,3}}

\institute{Dipartimento di Fisica, Universit\`{a} di Roma Tor Vergata, Via della Ricerca scientifica 1, I-00133 Roma, Italy
\and
Harvard, Smithsonian Astrophysical Observatory, 60 Garden Street, Cambridge, MA 02138, USA
\and
INAF , Via Fosso del Cavaliere 1, I-00100, Roma, Italy}

\offprints{alessandro.paggi@roma2.infn.it}
\date{Received April 1, 2009; accepted July 12, 2009}

\markboth{}
{}

\abstract
{The synchrotron-self Compton (SSC) radiation process is widely held to provide a close representation of the double peaked spectral energy distributions from BL Lac Objects (BL Lacs), which are marked by non-thermal beamed radiations, highly variable on timescales of days or less.
Their outbursts in the \(\gamma\) rays relative to the optical/X rays might be surmised to be enhanced in BL Lacs as these photons are upscattered via the inverse Compton (IC) process.}
{From the observed correlations among the spectral parameters during optical/X-ray variations we aim at predicting corresponding correlations in the \(\gamma\)-ray band, and the actual relations between the \(\gamma\)-ray and the X-ray variability consistent with the SSC emission process.}
{Starting from the homogeneous single-zone SSC source model, with log-parabolic energies distributions of emitting electron as required by the X-ray data of many sources, we find relations among spectral parameters of the IC radiation in both the Thomson (for Low energy BL Lacs) and the Klein-Nishina regimes (mainly for High energy BL Lacs); whence we compute how variability is driven by a smooth increase of key source parameters, primarily the root mean square electron energy.}
{In the Klein-Nishina regime the model predicts for HBLs lower inverse Compton fluxes relative to synchrotron, and milder \(\gamma\)-ray relative to X-ray variations. Stronger \(\gamma\)-ray flares observed in some HBLs like {\mrkb} are understood in terms of additional, smooth increases also of the emitting electron density.
However, episodes of rapid flares as recently reported at TeV energies are beyond the reach of the single component SSC model with one dominant varying parameter.
Furthermore, spectral correlations at variance with our predictions, as well as TeV emissions in LBL objects (like BL Lacertae itself) cannot be explained in terms of the simple HSZ SSC model, and in these cases the source may require additional electron populations in more elaborate structures like decelerated relativistic outflows or sub-jet scenarios.}
{We provide a comprehensive benchmark to straightforwardly gauge the capabilities and effectiveness of the SSC radiation process. The single component SSC source model in the Thomson regime turns out to be adequate for many LBL sources. In the mild Klein-Nishina regime it covers HBL sources undergoing variations driven by smooth increase of a number of source parameters.
However, the simple model meets its limits with the fast/strong flares recently reported for a few sources in the TeV range; these require sudden accelerations of emitting electrons in a second source component.}


\authorrunning{A. Paggi et al. }
\titlerunning{}

\maketitle
\section{Introduction}
Blazars are among the brightest Active Galactic Nuclei (AGNs) observed, with inferred (isotropic) luminosities up to \(L\sim {10}^{47}\mbox{ erg}\mbox{ s}^{-1}\). Actually these sources are widely held to be AGNs radiating from a relativistic jet closely aligned to the observer line of sight; this emits non-thermal radiations with observed fluxes enhanced by relativistic effects so as to often overwhelm other, thermal or reprocessed emissions.

BL Lacs in particular are Blazars showing no or just weak emission lines. Their spectra may be represented as a continuous spectral energy distribution (SED) \(S_\nu=\nu F_\nu\) marked by two peaks: a lower frequency one, widely interpreted as synchrotron emission by highly relativistic electrons; and a higher frequency counterpart, believed to be inverse Compton upscattering by the same electrons of seed photons either emitted by an external source (external Compton, EC, see e.g. Sikora et al. \cite{sikora}), or provided by the photons of the very synchrotron radiation (synchrotron-self Compton, SSC; see, e.g., Marscher \& Gear \cite{marscher1985}; Maraschi et al. \cite{maraschi}).

The BL Lacs are usually classified in terms of the frequency of their synchrotron peak (see Padovani \& Giommi \cite{padovani}): for the low frequency BL Lacs (LBLs) the peak lies in the infrared-optical bands, while the inverse Compton (IC) component peaks at MeV energies; instead, high frequency BL Lacs (HBLs) feature a first peak in the X-ray band and the second one at \(100\mbox{ GeV}\) energies or beyond; intermediate frequency BL Lacs (IBLs) stand in the middle, with the first peak at optical frequencies and the second peak at \(\sim 1 \mbox{ GeV}\).

BL Lac Objects also show rapid variability on timescales of days or shorter, during which the sources undergo strong flux variations often named ``flares''.

The homogeneous single zone (HSZ) SSC model (based, in particular, on a single electron population) constitutes an attractively simple source structure worth to be extensively tested and pushed to its very limits, as we discuss below; following Katarzy\'{n}ski et al. (\cite{katarzynsky}), Tramacere et al. (\cite{tramacere2007}), we focus on deriving the simultaneous changes of the synchrotron and IC fluxes of BL Lacs to be expected from an updated version of SSC. We use a realistic log-parabolic energy distribution of the electrons radiating the synchrotron photons and upscattering them by IC in either the Thomson or the Klein Nishina (KN) regime.

\section{Background relations}
\subsection{Log-parabolic electron distribution and spectra}\label{spectra}
The SEDs of BL Lacs exhibit two distinct peaks or humps. To describe these in detail a number of spectral models have been used over the years: power-laws with an exponential cutoff, broken power-laws, or spectra curved in the shape of a log-parabola. The latter are being increasingly used, as they generally provide more accurate fits than broken power-laws in the optical – X-ray bands, with residuals uniformly low throughout a wide energy range (see discussion by Landau et al. \cite{landau}; Tanihata et al. \cite{tanihata}; Massaro et al. \cite{massaro2004a}; Tramacere et al. \cite{tramacere2007}; Nieppola et al. \cite{nieppola}; Donnarumma et al. \cite{donnarumma})\footnote{The log-parabolic particle energy distribution could be made asymmetric by adding a powerlaw tail at low energies (see Massaro et al. \cite{massaro2006}) to represent also low frequency data, e. g., radio in addition to optical/X-ray data; however, here we will focus on the SEDs around their peaks and so we do not use such additions.}.

So we adopt log-parabolae in view of their optimal performance as a precise fitting tool. Moreover, they are also effective in yielding simple, analytical expressions (given in their exact form in Appendix \ref{calcoli}, following Eqs. \ref{curvaturasincro}, \ref{curvaturathomson} and \ref{curvaturakn}) for peak frequencies and fluxes in both in the inverse Compton regimes, Thomson and Klein-Nishina.
Finally, the physically interesting feature of log-parabolic spectra is their direct link to the acceleration processes of the emitting particles; in fact, as shown in Appendix \ref{accelerazione}, log-parabolic emitted spectra simply relate to log-parabolic particle energy distributions, that in turn naturally arise from systematic and stochastic acceleration processes as dealt with by the Fokker-Planck equation.

Thus for the synchrotron SED we write
\begin{equation}\label{logparsynch}
S^s_\nu = S^s_0\,{\left( {\frac{\nu}{\nu_0}}\right)}^{-(a_s-1)-b_s\log{\left( {\frac{\nu}{\nu_0}}\right) }};
\end{equation}
here \(a_s\) gives the constant component of the spectral index for the energy flux and \(b_s\) is the spectral curvature.
In the following we shall denote with \(\xi\) the frequency where the SED peaks and by \(S\) the related flux.
Spectral curvature values are observed to be around \(0.2\) (Massaro et al. \cite{massaro2004b}) and decreasing with time and/or frequency as given by Eqs. \ref{curvvstime} and \ref{curvvsxi} of Appendix \ref{accelerazione}; since the decrease is mild during flares (which are essentially dominated by systematic acceleration processes), we can  consider a nearly constant spectral curvature to a first approximation.
Then one finds the scaling relations \(\xi\propto B\,\gamma_p^2\,\delta\) and \(S\propto B^2\,\gamma_p^2\, n\, R^3\,\delta^4\), where \(n\) is the particle density, \(R\) is the emitting region radius, \(B\) is the magnetic field, \(\delta\) is the beaming factor, and \(\gamma_p\, m\, c^2\) is the square root of the mean quadratic energy (thereafter r.m.s. energy) of the particle distribution.

An analogous expression holds for the inverse Compton radiation
\begin{equation}\label{logparIC}
S^c_\nu\approx {S}^c_0\,{\left( {\frac{\nu}{\hat{\nu}_0}}\right)}^{-(a_c-1)-b_c\log{\left( {\frac{\nu}{\hat{\nu}_0}}\right) }},
\end{equation}
which in the Thomson regime the IC has \emph{less} curved spectra with \(b_c\approx b_s/2\). At nearly constant curvature
the parameters to focus on are \(\epsilon\propto B\,\gamma_p^4\,\delta\) and \(C\propto B^2\,\gamma_p^4\, n^2\, R^4\,\delta^4\) for the SED peak frequency and flux, respectively. In the extreme Klein-Nishina regime we obtain instead \emph{more} curved spectra with \(b_c\approx 5\, b_s\), and
\(\epsilon\propto \gamma_p\,\delta\) and \(C\propto B\, n^2\, R^4\,\delta^4\) hold (see Appendix \ref{calcoli} for more details, and Massaro et al. \cite{massaro2008a}, \cite{massaro2008b}).

\begin{figure}
 \centering
 \includegraphics[scale=0.35]{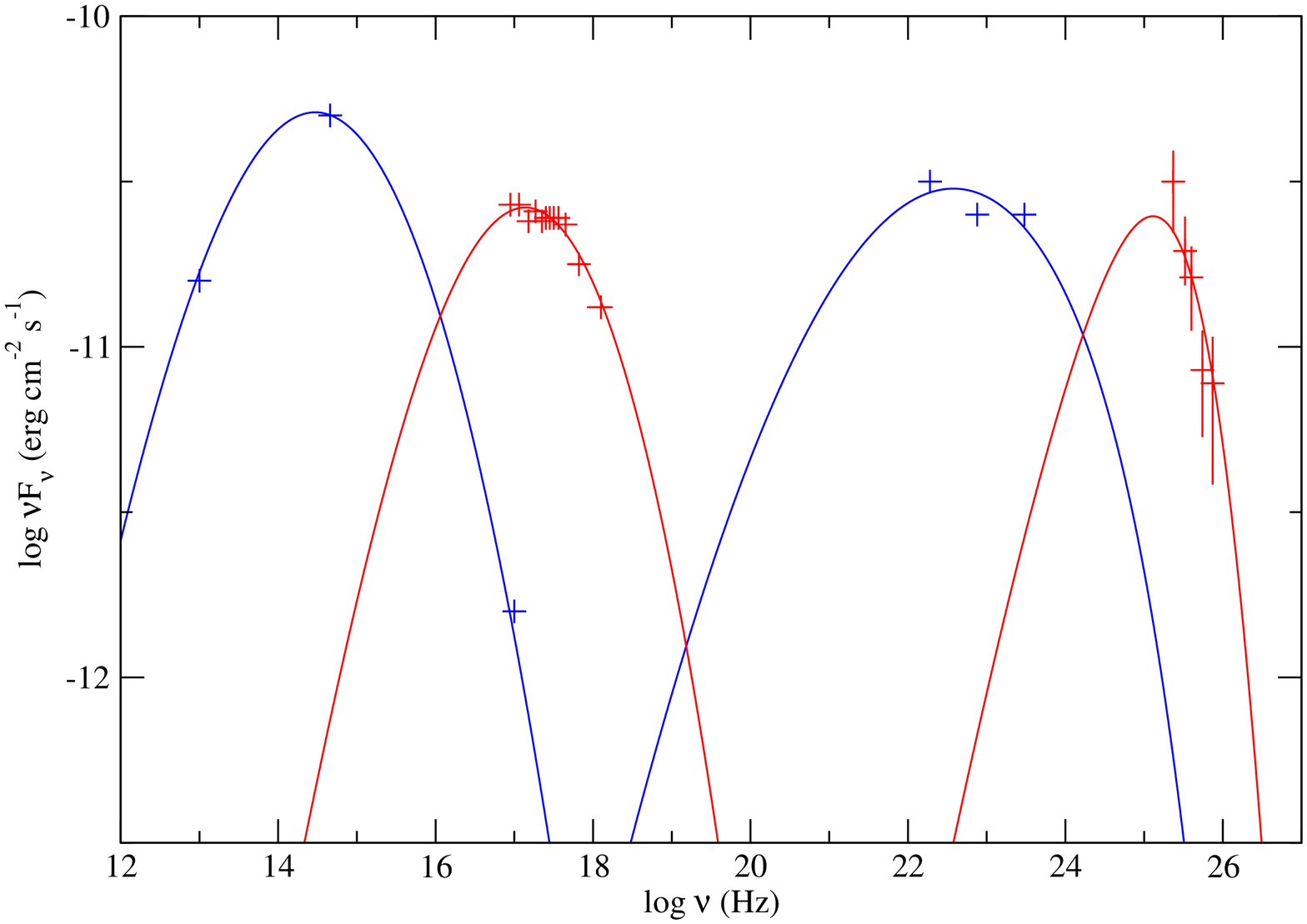}
\caption{Two examples of exact, steady SEDs provided by numerical simulations (code by Massaro \cite{massarotesi}) after the HSZ SSC model from relativistic electrons with a log-parabolic energy distribution \(N(\gamma)\propto \gamma^{-s-r\log{\gamma}}\) (see Appendix \ref{accelerazione}). The two characteristic humps correspond to synchrotron emission (left) and to inverse Compton scattering (right), either in the Thomson regime (blue line) or in the KN regime (red line). Standard sources that are so fitted include the low state of 0716+714 (an IBL, blue crosses) observations of November 1996 by {\sax}, AIT and EGRET (Tagliaferri et al. \cite{tagliaferri}); and 1ES 1553+113 (a HBL, red crosses) observations of April 2005 by {\swf} and April-May 2005 by MAGIC (Massaro \cite{massarotesi}).}
\label{esempio}
 \end{figure}

Such spectra are radiated by electron populations with a log-parabolic energy distribution, namely
\begin{equation}\label{logpar1}
N(\gamma)=N_0 {\left( {\frac{\gamma}{\gamma_0}}\right)}^{-s-r\log{\left( {\frac{\gamma}{\gamma_0}}\right) }}.
\end{equation}
The result is easily inferred and straightforward to compute on using the delta-function approximations for the single particle emissions (see Appendix \ref{calcoli}); it is found to hold also for the exact shape of the latter with the numerical simulations outlined in Fig. \ref{esempio}, and reported in full detail in Figs. \ref{evolthm} and \ref{evolkn}.

\subsection{Source parameters}\label{parametri}
Using the previous scalings, with the relation \(R= c\,\Delta t\,\delta/{(1+z)}\) between the observed variation time \(\Delta t\) and the size \(R\) of the emitting region (assuming spherical symmetry in the rest frame), we obtain relations of source parameters with the observables. In the Thomson regime we find
\begin{equation}\label{prima}
{\gamma_p}\propto
{\epsilon}^{\frac{1}{2}}{\xi}^{-\frac{1}{2}}
\end{equation}
\begin{equation}
{R}\propto
{\epsilon}^{\frac{1}{2}}\,{S}^{\frac{1}{2}}\,{\Delta t}^{\frac{1}{2}}\,{\xi}^{-1}\,{C}^{-\frac{1}{4}}
\end{equation}
\begin{equation}
{B}\propto
{\xi}^{3}\,{C}^{\frac{1}{4}}\,{\Delta t}^{\frac{1}{2}}\,{\epsilon}^{-\frac{3}{2}}\,{S}^{-\frac{1}{2}}
\end{equation}
\begin{equation}\label{deltathm}
{\delta}\propto
{\epsilon}^{\frac{1}{2}}\,{S}^{\frac{1}{2}}{\xi}^{-1}\,{C}^{-\frac{1}{4}}\,{\Delta t}^{-\frac{1}{2}}
\end{equation}
\begin{equation}\label{ultimathm}
{n}\propto
{\xi}^{2}\,{C}^{\frac{5}{4}}\, {\epsilon}^{-\frac{3}{2}}\,{S}^{-\frac{3}{2}}\,{\Delta t}^{-\frac{1}{2}}\, .
\end{equation}
For the extreme KN regime we find instead:
\begin{equation}\label{primakn}
{\gamma_p}\propto
{\xi}^{\frac{3}{5}}\,{\epsilon}^{\frac{3}{5}}\,{C}^{\frac{1}{5}}\,{\Delta t}^{\frac{2}{5}}{S}^{-\frac{2}{5}}
\end{equation}
\begin{equation}
{R}\propto
{\epsilon}^{\frac{2}{5}}\,{S}^{\frac{2}{5}}\,{\Delta t}^{\frac{3}{5}}\,{\xi}^{-\frac{3}{5}}\,{C}^{-\frac{1}{5}}
\end{equation}
\begin{equation}
{B}\propto
{\xi}^{\frac{2}{5}}\,{S}^{\frac{2}{5}}{\epsilon}^{-\frac{8}{5}}\,{C}^{-\frac{1}{5}}\,{\Delta t}^{-\frac{2}{5}}
\end{equation}
\begin{equation}\label{deltakn}
{\delta}\propto
{\epsilon}^{\frac{2}{5}}\,{S}^{\frac{2}{5}}\,{\xi}^{-\frac{3}{5}}\,{C}^{-\frac{1}{5}}\,{\Delta t}^{-\frac{2}{5}}
\end{equation}
\begin{equation}\label{ultima}
{n}\propto
{\xi}^{\frac{11}{5}}\,{C}^{\frac{7}{5}}\,{\epsilon}^{-\frac{4}{5}}\,{S}^{-\frac{9}{5}}\,{\Delta t}^{-\frac{1}{5}}\,.
\end{equation}
Note that a delay of the order of \(\Delta t\) may elapse between synchrotron and IC spectral variations.

Numerical values for the prefactors are given in Appendix \ref{constraints}, where normalizations are chosen referring to a typical LBL source with \(S\approx {10}^{-11}\mbox{ erg}\mbox{ cm}^{-2}\mbox{ s}^{-1}\), \(\xi\approx {10}^{14}\mbox{ Hz}\), \(C\approx {10}^{-11}\mbox{ erg}\mbox{ cm}^{-2}\mbox{ s}^{-1}\), \(\epsilon\approx {10}^{22}\mbox{ Hz}\); and to a typical HBL source with \(S\approx {10}^{-10}\mbox{ erg}\mbox{ cm}^{-2}\mbox{ s}^{-1}\), \(\xi\approx {10}^{18}\mbox{ Hz}\), \(C\approx {10}^{-11}\mbox{ erg}\mbox{ cm}^{-2}\mbox{ s}^{-1}\), \(\epsilon\approx {10}^{26}\mbox{ Hz}\). Values of physical parameters for some sources are given in Tab. \ref{tabella}. Note that the simple Eqs. (\ref{prima}) - (\ref{ultimathm}) provide parameter values in close agreement with the ones needed to fit the observations of LBL or IBL sources in Thomson regime; on the other hand the extreme KN limit underlying Eqs. (\ref{primakn}) - (\ref{ultima}) would yield disagreement for HBLs, indicating that these sources just border the KN regime.

\begin{table*}[ht]
\centering
\begin{tabular}{|c|c|c|c|c|c|c|c|}
\hline
Source name & \(z\) &\(\gamma_p\) & \(R\,(\mbox{cm})\) & \(B\,(\mbox{G})\) & \(\delta\) & \(n\,(\mbox{cm}^{-3})\)& \(r\)
\\
\hline
0716+714 (low state) & \(0.31\) & \(3.36\,{10}^3\) & \(2.61\,{10}^{17}\) & \(1.51\,{10}^{-1}\) & \(10\) & \(0.41\) & \(1.18\)
\\
0716+714 (high state) & ,, & \(9.92\,{10}^3\) & \(7.79\,{10}^{17}\) & \(7.64\,{10}^{-2}\) & \(10\) & \(0.03\) & \(1.18\)
\\
1ES 1553+113  & \(0.25\) & \(3.39\,{10}^4\) & \(0.64\,{10}^{16}\) & \(9.80\,{10}^{-1}\) & \(10\) & \(2.04\) & \(1.49\)
\\
\mrkb (low state)  & \(0.034\) & \(1.88\,{10}^5\) & \(0.38\,{10}^{16}\) & \(9.29\,{10}^{-2}\) & \(15\) & \(0.62\) & \(0.81\)
\\
\mrkb (high state) & ,, & \(2.58\,{10}^5\) & \(0.59\,{10}^{15}\) & \(3.85\,{10}^{-1}\) & \(15\) & \(23.49\) & \(0.74\)
\\
\hline
\end{tabular}
\caption{Source parameters for some BL Lac sources.}
\label{tabella}
\end{table*}

We evaluate the position \(\xi_T\) of the synchrotron peak when the transition between the two IC regimes occurs, to find
\begin{equation}\label{passaggiogamma}
\xi_T\approx 7.15\,{10}^{15}\,{\left({\frac{B}{0.1\mbox{ G}}}\right)}^{\frac{1}{3}}\,
{\left({\frac{\delta}{10}}\right)}\,(1+z)^{-1}\mbox{ Hz}\, ,
\end{equation}
or equivalently
\begin{equation}\label{passaggiob}
\xi_T\approx 1.96\,{10}^{16}\,{\left({\frac{\gamma_p}{{10}^4}}\right)}^{-1}\,
{10}^{\frac{1}{2}\left({1-\frac{1}{5b}}\right)}\,
{\left({\frac{\delta}{10}}\right)}\,(1+z)^{-1}\mbox{ Hz}\, .
\end{equation}
So we conclude that in LBLs, which feature the synchrotron peak at optical/IR frequencies with \(\xi\approx {10}^{14}\mbox{ Hz}\), the majority of photons of the IC peak are upscattered in the Thomson regime; whereas in extreme HBLs, which show the synchrotron peak in the X-ray band with \(\xi\approx {10}^{18}\mbox{ Hz}\), the majority of IC peak photons are upscattered in the Klein-Nishina regime.

\subsection{Spectral correlations during X-ray spectral variations}
Correlations in the synchrotron emission between \(S\) and \(b_s\) with \(\xi\) can be used to pinpoint the main \emph{driver} of the spectral changes during X-ray variations (see also Tramacere et al. \cite{tramacere2007}). In fact, the synchrotron SED peak scales (for nearly constant spectral curvature as anticipated in \S\ref{spectra})
as \(S \propto n\,R^3\, \gamma_p^2\, B^2\, \delta^4\) at the peak frequency itself scaling as \(\xi \propto \gamma_p^2\, B\, \delta\).

Thus on writing the dependence of \(S\) on \(\xi\) in the form of a powerlaw \(S\propto \xi^\alpha\), we expect \(\alpha = 1\) to apply when the spectral changes are driven mainly by variations of the electrons r.m.s. energy; \(\alpha = 2\) for dominant changes of the magnetic field; \(\alpha = 4\) if changes in the beaming factor dominate; \(\alpha = \infty\) formally applies for changes only in the number density of the emitting particles. Results by the latter authors focus on \(\gamma_p\) and \(B\) as dominant drivers.
Starting from such correlations we aim at predicting the expected correlations and flux variations in the \(\gamma\)-ray band.

Particle r.m.s. energy variations are of particular interest because they naturally arise as a consequence of systematic plus stochastic acceleration of the electron population (details in Appendix \ref{accelerazione}).

\section{\(\gamma\)-ray spectra}

\subsection{Spectral correlations}\label{correlazioni}

The above results lead us to expect specific correlations between SED peak frequencies and fluxes as follows. In the case of dominant r.m.s. particle energy variations we expect for the synchrotron emission
\begin{equation}\label{variazprima}
 S\propto \xi\, ,
\end{equation}
 while for the IC we have
\begin{equation}
 C\propto \epsilon
\end{equation}
for LBL objects, and
\begin{equation}
 C\approx {const}
\end{equation}
for extreme HBLs.

In the case of magnetic field variations we have for the synchrotron emission
\begin{equation}
 S\propto \xi^2\, ,
\end{equation}
while for the IC we have
\begin{equation}
C\propto \epsilon^2
\end{equation}
for LBL objects, and
\begin{equation}\label{variazultima}
 C\propto \epsilon^\tau
\end{equation}
formally with \(\tau=\infty\) for extreme HBLs.

To complete the above picture, we consider how saturation affects correlations. Denoting the total number of particles with \(N_T\), in the jet frame we have for the synchrotron luminosity \(L_s \propto N_T\,\gamma_p^2\), while for IC luminosity we have \(L_C \propto {\left( {N_T\,\gamma_p^2}\right)}^2\) and \(L_C \propto N_T^2\) in Thomson and KN regime, respectively. Consider the case where the total power available to the jet is limited, e. g., in the BZ mechanism (Blandfor \& Znajek \cite{bz}) of power extraction from a maximally rotating black hole with an extreme, radiation pressure dominated disk that holds the pressure of the magnetic field facing the black hole horizon (see discussion by Cavaliere \& D'Elia \cite{cavaliere}, and references therein). Then the power has an upper bound \(L_{BZ}\approx 2\,{10}^{45} M_9\mbox{ erg}\mbox{ s}^{-1}\); as a consequence, we expect an effect of saturation on the SED peaks as the emitted power, which represents a substantial fraction of the total jet power (see Celotti \& Ghisellini \cite{celotti}), approaches \(L_{BZ}\). For the total power, including radiative, kinetic and magnetic components, \(L_{tot}\approx L_{BZ}\) we envisage \(N_T\sim\gamma_p^{-2}\), that is, the particles effectively accelerated to the increasing \(\gamma_p\) decrease in number on approaching the BZ limit.

The data analyses by Tramacere et al. (\cite{tramacere2007}, \cite{tramacere2009}) for the variations of {\mrka} are consistent with the beginning of such a saturation effect, as they show a declining value of \(\alpha\) when the peak moves toward the highest energies and approaches \(L\approx {10}^{46}\mbox{ erg}\mbox{ s}^{-1}\). Partial saturation effects can also be explained in terms of systematic and stochastic acceleration processes; in particular, when the stochastic prevails over the systematic acceleration (as is bound to occur at high energies, see Appendix \ref{calcoli}) we expect \(\alpha\) to decrease toward \(0.6\), i.e. \(S\propto \xi^{0.6}\).

\subsection{Related variabilities}
We now focus on related variabilities between IC and synchrotron fluxes; here we are mainly interested on increases of peak frequencies and fluxes, as driven by acceleration processes of the emitting particles or by a growing magnetic field (see also Katarzy\'{n}ski et al. \cite{katarzynsky}); in Appendix \ref{newappendix} we briefly discuss the problems posed by the decay phase as stressed by Katarzy\'{n}ski et al.

\begin{figure}
 \centering
 \includegraphics[scale=0.35]{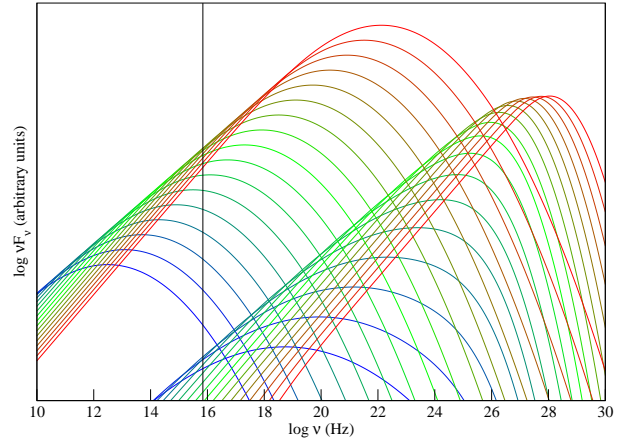}
\caption{To illustrate the differences of the IC SEDs in the Thomson and in the extreme KN regime, we show the results of numerical simulations (code by Massaro \cite{massarotesi}) after the HSZ SSC model, for different sources with larger and larger r.m.s. energy of the radiating particles and other parameters kept constant (including the curvature of the particle energy distribution). The vertical line indicates the frequency of the synchrotron peak where the IC scattering changes over the Thomson to the KN regime (see Eq. (\ref{passaggiogamma})). Actual sources in their evolution (represented in Fig. \ref{evolthm} and \ref{evolkn}) vary their spectral curvature somewhat and only span a limited range in frequency and flux to the left (LBLs) or to the right (HBLs) of the line, with only the IBLs likely to cross it.}
\label{saturgamma}
 \end{figure}

\begin{figure}
 \centering
 \includegraphics[scale=0.35]{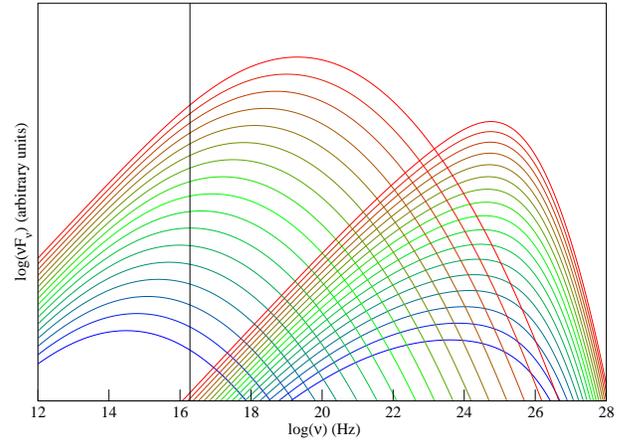}
\caption{As in Fig. \ref{saturgamma}, we show SEDs obtained from numerical simulations (code by Massaro \cite{massarotesi}) of HSZ SSC radiations, for different sources with larger and larger magnetic field \(B\) and  other parameters kept constant. The vertical line again indicates the transition frequency from the Thomson to the KN regime (see Eq. (\ref{passaggiob})); as \(B\) increases, the IC scattering changes its regime, and reduced cross section yields similar compression effects as in Fig. \ref{saturgamma}.}
\label{saturb}
 \end{figure}

For LBLs, that radiate mostly in the Thomson regime, we predict from Eqs. (\ref{variazprima}) - (\ref{variazultima}) a quadratic or linear variation of \(C\) with respect to \(S\), depending on the parameter that mainly drives source variations. That is to say, we expect
\begin{equation}\label{variazgammaS}
{\Delta \ln C}\approx 2\, {\Delta \ln S}
\end{equation}
in the case of dominant particle r.m.s. energy variations, or
\begin{equation}\label{variazbS}
{\Delta \ln C}\approx  {\Delta \ln S}
\end{equation}
for dominant magnetic field variations (see also Figs. \ref{saturgamma} and \ref{saturb}).

Instead, for HBLs that radiate closer to the Klein-Nishina regime, we obtain a weaker \(\gamma\)-ray variability due to the decreasing KN cross section, leading in the extreme to
\begin{equation}\label{variazgammaC}
{\Delta \ln C}\approx 0
\end{equation}
in case of dominant particle r.m.s. energy variations, or
\begin{equation}\label{variazbC}
{\Delta \ln C}\approx  \frac{1}{2}\,{\Delta \ln S}
\end{equation}
for dominant magnetic field variations. Expected correlations are summarized in Table \ref{tabella2}.

From Figs. \ref{saturgamma} and \ref{saturb} it is seen that the transition to the KN regime has three effects on the IC spectrum: first, it \emph{brakes} the peak frequency increase, because the energy the photon can gain is limited to the total electron energy \(\gamma\, m\,c^2\); second, it \emph{reduces} the flux increase at the SED peak, as a consequence of reduced cross section; third, it \emph{increases} the spectral curvature near the peak as a consequence of frequency compression.

Viceversa, variations observed in IC section of the spectrum are related via Eqs. (\ref{variazgammaS}) - (\ref{variazbC}) to variations expected in the synchrotron emission; in particular, in HBLs \(\gamma\)-ray variations ought to have enhanced counterparts in X rays.

Up to now Figs. \ref{saturgamma} and \ref{saturb} may be interpreted as a collection of different sources or of unrelated states of a single source; henceforth we will focus on the interpretation in terms of \emph{evolution} in a flaring source where the parameters vary in a continuous fashion; in particular, \(\gamma_p\) increases under the drive of systematic and stochastic electron accelerations, in keeping with the description in terms of a continuity or kinetic equation of the Fokker-Planck type as given in detail in Appendix \ref{accelerazione}. Signatures of such an evolution are provided not only by continuous growth of the peak frequencies, but even more definitely by the \emph{irreversible} decrease of the spectral curvature (in terms of time or frequency) under the drive of the stochastic acceleration, see Eq. (\ref{curvvsxi}) and the observations by Tramacere et al. (\cite{tramacere2007}, \cite{tramacere2009}). In fact, for a single electron population the curvature is to slowly decrease or stay nearly constant on the timescale of the systematic acceleration. So a sudden increase of curvature signals the injection of a new population/component.

In the following we will neglect radiative cooling, which does not strongly affect the spectral shape around the peaks; such a process will be particularly relevant when the source, after flaring up, relaxes back to a lower state upon radiating away its excess energies from high frequencies downwards (see Appendix \ref{accelerazione}).

\begin{table*}[ht]
\centering
\begin{tabular}{|l|l|c|c|}
\hline
process & peak flux and frequency & flux-frequency correlation & \(S-C\) correlation\\
\hline
synchrotron & \(\begin{array}{l}
                S\propto R^3\,B^2\,\gamma_p^2\, n\, \delta^4 \\
                \xi\propto B\,\gamma_p^2\,\delta
              \end{array}\) &\(S\propto\xi^\alpha\left\{{\begin{array}{c}
                \alpha=1 \,\left({\gamma_p}\right) \\
                \alpha=2 \,\left({B}\right)
              \end{array}}\right.\)&\\
\hline
IC Thomson & \(\begin{array}{l}
                C\propto R^4\,B^2\,\gamma_p^4\, n^2\, \delta^4 \\
                \epsilon\propto B\,\gamma_p^4\,\delta
              \end{array}\) &\(C\propto\epsilon^\alpha\left\{{\begin{array}{c}
                \alpha=1 \,\left({\gamma_p}\right) \\
                \alpha=2 \,\left({B}\right)
              \end{array}}\right.\)& \(\begin{array}{l}
                C\propto S^2 \\
                C\propto S
              \end{array}\)\\
\hline
IC KN & \(\begin{array}{l}
                C\propto R^4\,B\, n^2\, \delta^4 \\
                \epsilon\propto \gamma_p\,\delta
              \end{array}\) &\(C\propto\epsilon^\alpha\left\{{\begin{array}{c}
                \alpha=0 \,\left({\gamma_p}\right) \\
                \alpha=\infty \,\left({B}\right)
              \end{array}}\right.\)&\(\begin{array}{l}
                C\approx{const}\\
                C\propto S^{1/2}
                \end{array}\)\\
\hline
\end{tabular}
\caption{Spectral correlations (for \(r\backsimeq const\)). We have denoted with \((\gamma_p)\) or \((B)\) the variations driven by increases of rms electron energy or magnetic field, respectively.}
\label{tabella2}
\end{table*}

\begin{figure}
 \centering
 \includegraphics[scale=0.35]{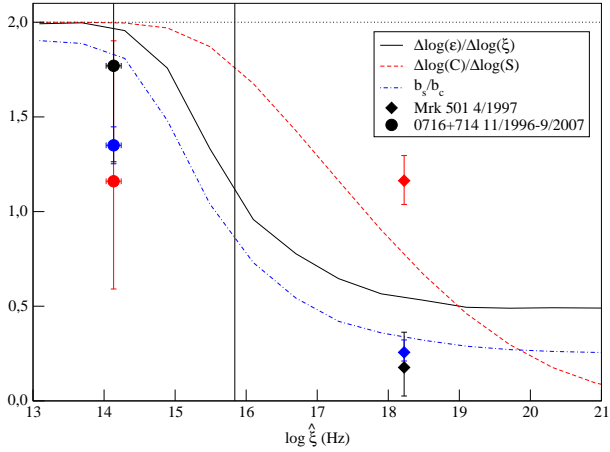}
\caption{Computed changes of spectral parameters corresponding to extended increase of r.m.s. particle energy with the other source parameters kept constant. As functions of \(\hat\xi=\xi\,{(B/0.1\mbox{ G})}^{-1/3}{(\delta/10)}^{-1}(1+z)\) (where \(\xi\) is the synchrotron SED peak frequency, see Eq. (\ref{passaggiogamma})), the full line represents the ratio \(\Delta\ln{\epsilon}/\Delta\ln{\xi}\), the dashed line the ratio \(\Delta\ln{C}/\Delta\ln{S}\) and the dot-dashed line the curvature ratio \(b_s/b_c\) at the peaks. Circles represent the parameters of 0716+714 as observed on November 1996 and September 2007 (Tagliaferri et al. \cite{tagliaferri}; Chen et al. \cite{chen}), and diamonds represent {\mrkb} observations of April 1997 (Djannati-Atai et al. \cite{djannati}; Massaro et al. \cite{massaro2006}), each with the same colours as the lines. Changes of \(n\) only would yield \(\Delta\ln{C}=2\,\Delta\ln{S}\) at all frequencies, represented by the horizontal dotted line; added to a pure r.m.s. energy increase, a moderate increase of \(n\) can account for the upward deviations of observed values.}
\label{pargamma}
 \end{figure}

\begin{figure}
 \centering
 \includegraphics[scale=0.35]{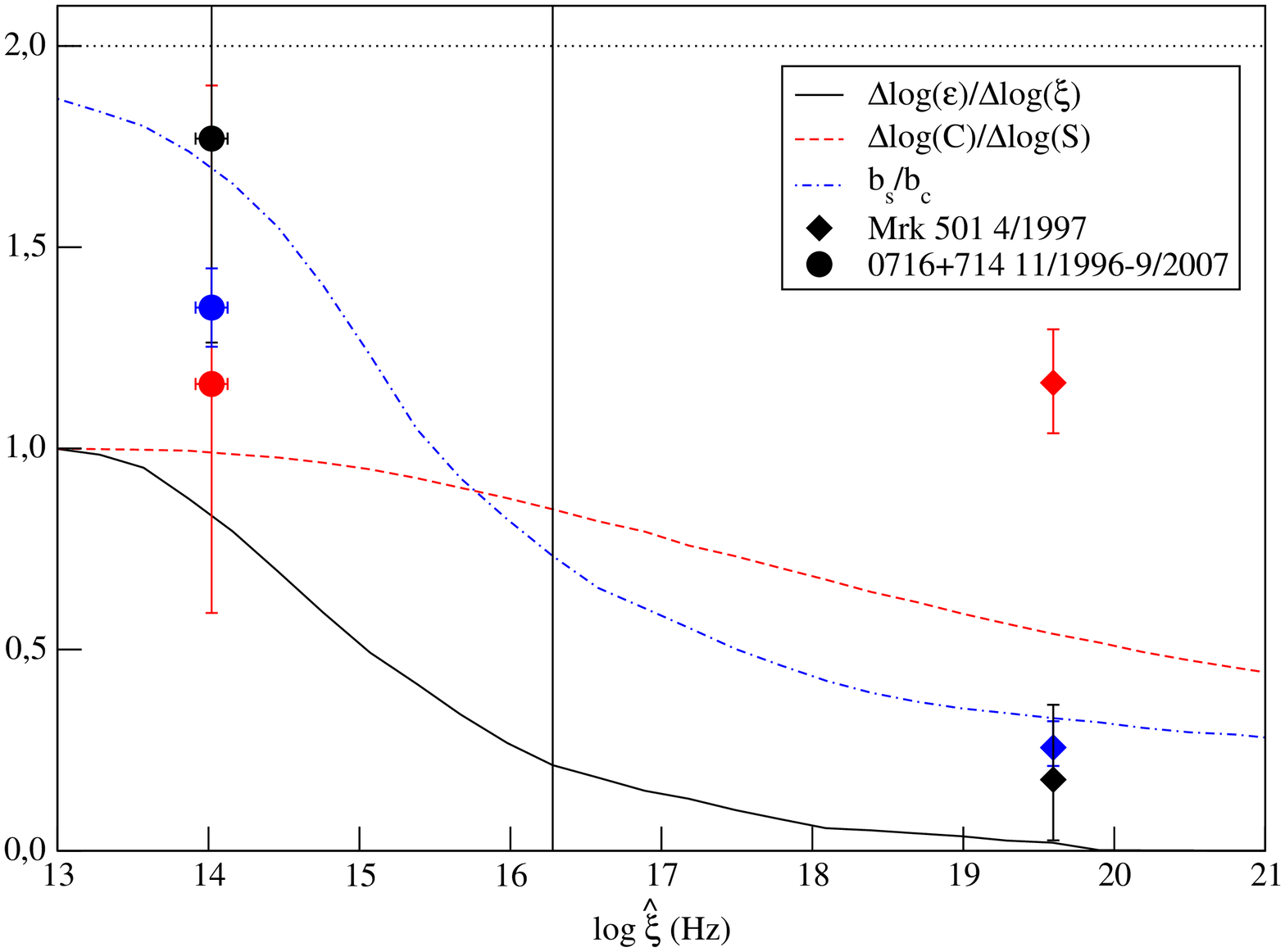}
\caption{Computed changes of spectral parameters as driven by an extended increase of the magnetic field with the other source parameters kept constant. The black full line represents the value of the ratio \(\Delta\ln{\epsilon}/\Delta\ln{\xi}\), the red dashed line represents the value of the ratio \(\Delta\ln{C}/\Delta\ln{S}\) and the blue dot-dashed line represents the ratio \(b_s/b_c\), for different values of \(\hat\xi=\xi\,{(\gamma_p/{10}^4)}{10}^{(1-1/5b)/2}{(\delta/10)}^{-1}(1+z)\) (where \(\xi\) is the synchrotron SED peak frequency, see Eq. (\ref{passaggiob})), data points are the same as in Fig. \ref{pargamma}; compared with the latter considerably larger changes of \(n\) are needed to recover the observed values.}
\label{parb}
 \end{figure}

\subsection{Specific sources}\label{sorgenti}

We have applied the above simple expectations to two sources, 0716+714 (IBL) and {\mrkb} (HBL), among the few to date to provide extended simultaneous coverage of two different source states in X rays and \(\gamma\) rays.

\begin{figure}
 \centering
 \includegraphics[scale=0.35]{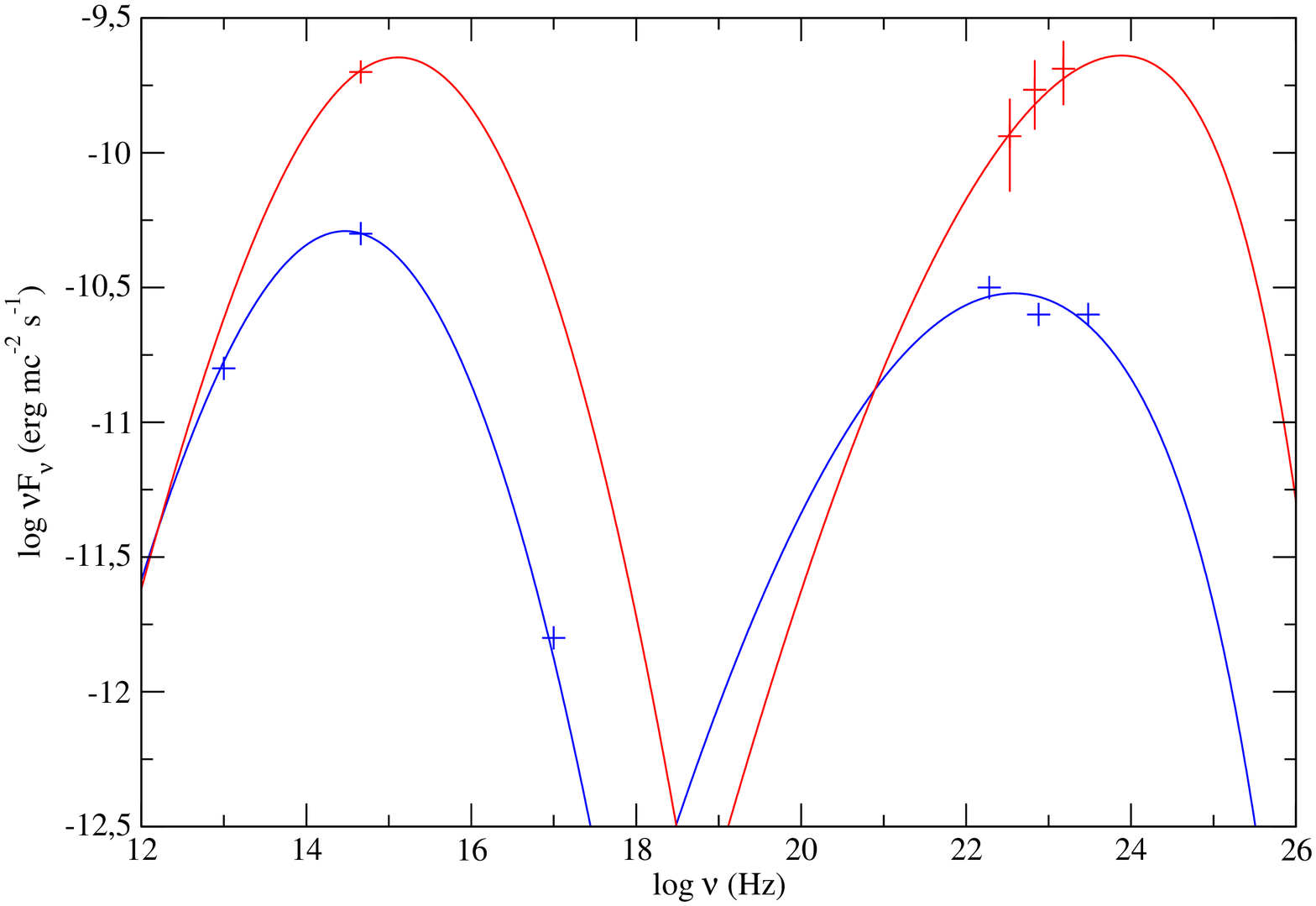}
\caption{Spectral fit of two different states of the IBL source 0716+714. The low state (blue line) data are relative to 1996 November 14 simultaneous observation by AIT, {\sax} and {\egret}; the high state (red line) data are relative to 2007 September 7-12 simultaneous observations by GASP-WEBT and {\agile}-GRID (Tagliaferri et al. \cite{tagliaferri}; Chen et al. \cite{chen}).}
\label{0716}
 \end{figure}

The first is widely considered to be an IBL. In the low state observed by Automatic Imaging Telescope (AIT) in November 14 1996, the synchrotron peak at about at about \(10^{14.5}\mbox{ Hz}\) falls quite below the threshold frequency expressed by Eqs. (\ref{passaggiogamma}) or (\ref{passaggiob}), so we expect for this source the IC scattering to occur mostly in the Thomson regime; simultaneous \(\gamma\)-ray observations were carried out by {\egret}. The high activity
state of the source is described by data relative to GASP project of the WEBT and \agile-GRID observations of 2007 September 7-12 (see Fig. \ref{0716}). The source has shown similarly low flux levels during EGRET observations \cite{tagliaferri}; so for flares with short duty cycle, it is not unreasonable to evaluate variations between the two distant states in the absence of closer observations; curvature variation is not required to fit the data, as we expect at lower energies where systematic dominates the stochastic acceleration (see Appendix \ref{calcoli}).
As shown in Figs. \ref{pargamma} and \ref{parb}, peak flux and frequency variations of 0716+714 appear to be in between the lines representing increasing r.m.s. particle energy and increasing magnetic field, and so they may be described in terms of HSZ SSC with simultaneous variations of these two parameters.

For the HBL source {\mrkb} we consider the two states described by Massaro et al. (\cite{massaro2006}), with simultaneous {\sax} and CAT observations of April 7 and 16 1997.
To understand the behavior of this source, from Figs. \ref{pargamma} and \ref{parb} it is seen that variations of \(B\) alone are ineffective and require complementary \(n\) increase by a factor of about \(30\); variations of \(\gamma_p\) are adequate in the KN regime, and only require a complementary doubling of \(n\) (which by itself would yield \({\Delta \ln C}\approx  2\,{\Delta \ln S}\) in both scattering regimes) to yield higher \(\gamma\)-ray flux increase. Note that in going to higher energies the spectral curvature \textit{decreases} somewhat (from \(b=0.161\pm 0.007\) to \(b=0.148\pm 0.005\) for the low and high state, respectively) in keeping with model predictions; such a behavior is interesting as it marks a \emph{smooth} growth, if anything, in the number of emitting particles as contrasted with sudden, substantial re-injection of nearly monoenergetic electrons that would suddenly \textit{reverse} the otherwise \emph{irreversible} decrease of \(r\) and \(b\). Such a smooth growth may obtains in a scenario of an expanding blast-wave that progressively involves more electrons (see, e. g., Ostriker \& McKee \cite{ostriker}; Lapi et al. \cite{lapi}; Vietri \cite{vietri}).

We stress that in HBL sources even small flux variations in \(\gamma\) rays are expected to have enhanced and observable counterparts in X rays. We suggest that such variations should be checked upon \(\gamma\)-ray "alarms", the inverse triggering relative to usual.
The absence of such lower energies counterparts will indicate, for example, a flare driven dominantly by a particle number density increase associated with a magnetic field decrease \(n\propto {B}^{-2}\), causing \(S\approx const\) but a decreasing synchrotron peak frequency; such a kind of flare may easily drown into the primary synchrotron component.

\subsection{Limiting timescales}\label{timescales}

As shown in Appendix \ref{accelerazione}, systematic and stochastic acceleration processes occur in the sources on timescales \(t_1=1/\lambda_1\) and \(t_2=1/\lambda_2\), respectively.
Limits to the variability timescales \(\Delta t\) relate to the apparent size of the emission region by \(\Delta t\geqslant {R}\,(1+z)/{c\,\delta}\) (see \S  \ref{parametri}); rapid variations require small emitting sources, but these could be optically thick to the pair production process. In fact, \(\gamma\)-ray photons collide with less energetic ones to produce \(e^\pm\) pairs, and the cross section of this process tops at a value around \(\sigma_T/5\) (where \(\sigma_T\) is the Thomson cross section) when the \(\gamma\)-ray photon frequency \(\nu\) and the target photons frequency \(\nu_t\) satisfy the relation \(\nu_t\approx {{m^2\,c^4\,\delta^2}}/{\left[{h^2\,{(1+z)}^2\,\nu}\right]}\).

For the \(\gamma\) rays to escape from the emitting region the source "compactness" intervenes and can set lower bounds on the beaming factor (Cavaliere \& Morrison \cite{cavmor}; Massaro \cite{massarotesi}; Begelman, Fabian \& Rees \cite{bfr}). In fact the optical depth for this process is expressed in terms of observational quantities (primed quantities refer to the jet rest frame) as
\begin{equation}
 \tau_{\gamma\gamma} \left({\nu_\gamma '}\right)=R\,\frac{\sigma_T}{5}\,n'_{ph}\left({\nu'_t}\right)\approx \frac{9}{20}\frac{\sigma_T\,h}{m^2\,c^6}\,
\frac{D^2}{\delta^6}\,{(1+z)}^4\,\frac{S\,\nu}{\Delta t}\,{\left( {\frac{\nu_t}{\xi}}\right)}^{-\beta}\, ,
\end{equation}
where \(D\) is the source distance expressed in \(\mbox{Gpc}\), with the shorthand \(\beta\equiv b\log{\left( {\frac{\nu_t}{\xi}}\right)}\). Escaping \(\gamma\)-ray radiation requires the source to be optically thin, that is, \( \tau_{\gamma\gamma} \left({\nu_\gamma '}\right)\leq 1\), resulting in
\begin{equation}
\delta\geq {\left[{\frac{9}{20}\frac{\sigma_T\,h^{1+2\beta}}{m^{2+2\beta}\,c^{6+4\beta}}\,
D^2\,{(1+z)}^{4+2\beta}\,\frac{S\,\nu^{1+\beta}\,\xi^\beta}{\Delta t}}\right]}^{\frac{1}{6+2\beta}}.
\end{equation}
Thus rapid flux variations set a lower bound on the beaming factor; the highest bound between the previous relation and \(\delta \geqslant {R}\,(1+z)/{c\,\Delta t}\) is to be relevant.

The exponent \(\beta\) contains a weak logarithmic dependence on \(\delta\) that may be neglected to a first approximation, and has a magnitude which depends on the source spectral properties; for definiteness, we use for \(b\) the typical value \(0.2\), and for \(\nu\approx {10}^{27}\mbox{ Hz}\). For a LBL with \(\xi\sim {10}^{14}\mbox{ Hz}\) we obtain \(\beta\approx 0.4\), while for an extreme HBL with \(\xi\sim {10}^{18}\mbox{ Hz}\), we obtain \(\beta\approx -0.5\). A variation timescale \(\Delta t\sim 5\) minutes for a LBL with \(S\sim 10^{-11}\mbox{ erg}\mbox{ cm}^{-2}\mbox{ s}^{-1}\) implies \(\delta\geqslant 30\), while in an extreme HBL with \(S\sim 10^{-10}\mbox{ erg}\mbox{ cm}^{-2}\mbox{ s}^{-1}\) we find \(\delta\geqslant 15\).

In the specific case of PKS 2155-304 discussed by Begelman et al. (\cite{bfr}), the emitted power \(L\approx {10}^{46}\mbox{ erg}\mbox{ s}^{-1}\) corresponds to \(S\sim 4\,{10}^{-11}\mbox{ erg}\mbox{ cm}^{-2}\mbox{ s}^{-1}\), with \(\xi\sim {10}^{16}\mbox{ Hz}\), to give \(\beta\approx 0.1\) and \(\delta\geqslant 50\) for \(\Delta t\sim 5\) minutes. We note that the source, even if widely considered a HBL, shows peculiar behaviors with respect to other TeV HBLs (see Massaro et al. \cite{massaro2008b}); moreover, it does not satisfy the condition \(\xi\gg\xi_T\) and so it cannot be considered an extreme HBL, as discussed below.

Extreme values of the beaming factor \(\delta\sim 50\) have been proposed to account for large peak separations (Konopelko et al. \cite{konopleko}) or to explain very rapid spectral variations (Begelman, Fabian \& Rees \cite{bfr}), formally with \(\Delta t \sim 5\) minutes for the \(\gamma\)-ray flares of PKS 2155-304 in 2006 July 28 (Aharonian et al. \cite{aharonian2007}) and {\mrkb} in June 30 and July 9 2007 (Albert et al. \cite{albert2007a}). Note that with BL Lac spectra realistically curved, the observed flux variations may be enhanced due to a slope effect best legible on the differential flux \(F_\nu\); that is, when fluxes are measured at frequencies where the spectral slope is steep (as may be the case for PKS 2155-304), a strong observed flux variation implies only a mild variation of the peak flux and requires smaller values of \(\delta\). Otherwise, a peak variation on a scale \(\Delta t \sim 5\) minutes would require in the SSC model \(\delta \sim 100\) for LBLs where \(\delta\propto {\Delta t} ^{-\frac{1}{2}}\) (see Eq. (\ref{deltathm})), and \(\delta \sim 20\) for an extreme HBL where \(\delta\propto {\Delta t} ^{-\frac{2}{5}}\) (see Eq. (\ref{deltakn})).

As stated under \S \ref{sorgenti}, in HBL sources even small flux variations in \(\gamma\) rays should have enhanced counterparts in X rays. These may become hard to observe for example in a flare dominantly driven by particle number density increase associated with a magnetic field decrease (i.g., with \(n\propto B^{-2}\)), leading to a synchrotron emission easily drowned into other components.
On the other hand, components with \(\delta > 20\) appearing in an HBL spectrum go beyond the HSZ SSC model with variations of one dominant parameter, and therefore require a more elaborate source structure.

In such cases the next natural scenarios are provided by decelerated relativistic outflows (Georganopoulos \& Kazanas \cite{gk2003}), or by nested spine-layer jets (Tavecchio \& Ghisellini \cite{tavecchio2008}) and  jets in a jet (Giannios et al. \cite{giannios}).

\section{Beyond the SSC}

In Table \ref{tabella2} we have provided a comprehensive \emph{benchmark} to gauge the performance of the HSZ SSC model for flaring BL Lac objects. This allows us to easily recognize events that may be accounted for within the model with physical variations of key source parameters, or that instead require more complex source structures.

Toward that purpose, we have used realistic log-parabolic spectral shapes produced by log-parabolic energy distributions of the emitting electrons, and studied IC radiation both in the Thomson and the Klein-Nishina regimes.

In the model we expect \(S\) to dominate \(C\) fluxes; in fact, moving to higher energies in a collection of different sources or in a prolonged evolution of a given source, we expect the emission to drift out of the pure Thomson and approach the KN regime, where the cross section decreases and limits the IC fluxes enforcing \(C\lesssim S\). Such a relation is found to hold in a number of sources, and in particular in HBLs (see for example Tagliaferri et al. \cite{tagliaferri08}).

From Table \ref{tabella2} we stress here the source spectral \emph{variations} predicted in \(\gamma\) rays. For LBLs, where the IC scattering mostly occurs in the Thomson regime, we recover the standard quadratic increase in \(\gamma\)-ray fluxes with respect to IR-optical ones, expressed by \({\Delta \ln C}\approx 2\, {\Delta \ln S}\) for dominant particle r.m.s. energy variations (see Eq. \ref{variazgammaS}). Instead for HBLs, in which the IC scattering approaches the Klein-Nishina regime, we expect \emph{smaller} or even vanishing increases of the \(\gamma\)-ray flux relative to X rays, that is, \({\Delta \ln C}\backsimeq 0\) for dominant particle r.m.s. energy variations (see Eq. \ref{variazgammaC}).

Comparing with specific sources, we find the HSZ SSC model to be adequate for most LBLs at the present observational levels; whilst for example the HBL source {\mrkb} in April 1997 showed a \(\gamma\)-ray flux increase appreciably stronger than expected if it were dominated by just one driving parameter. We explain this behavior within the model in terms of an additional, \textit{smooth} and moderate increase (by a factor around \(2\)) in number density of the electrons responsible for the emission (see \S \ref{sorgenti}).
Such conditions can still be provided by a \emph{single} electron population; this should be marked by a continuously \textit{decreasing} spectral curvature \(b\), as indicated by current data (Massaro et al. \cite{massaro2006}), a feature providing a potentially powerful signature to closely check on further data. Instead, for dominant magnetic field variations we would expect \({\Delta \ln C}\approx  {\Delta \ln S}\) for LBL sources (see Eq. (\ref{variazbS}), and \({\Delta \ln C}\approx \frac{1}{2}{\Delta \ln S}\) for HBL sources (see Eq. \ref{variazbC}); but in that case the required increase in particle number density should be much \emph{higher} by a factor of about \(30\), hard to interpret in terms of a single electron population.

On the other hand, on using the previous relations inversely, we see that even small \(\gamma\)-ray variations in HBL sources ought to have \emph{enhanced} counterparts in the X rays, unless flare activity were driven by an additional jet component, for example one with higher magnetic field but lower particle density; thus the corresponding emission is easily overwhelmed by, or drowned into the main synchrotron, but then the rest frame acceleration times must be short enough to involve a significant fraction of the emitting region. We suggest this is as a critical test for the simple model.

Another limitation arises from \emph{rapid} peak flux variations requiring large values of \(\delta\) as shown in \S \ref{timescales}. In the extreme, the few sources with particularly fast \(\gamma\)-ray increases observed so far, like PKS 2155-304 in 2006 July 28 and {\mrkb} in June 30 and July 9 2007 (see \S \ref{timescales}), require \emph{additional} components with very high beaming factors that go definitely beyond the simple SSC model.

Finally, another problem for the model arises from LBL objects showing substantial emissions in the GeV-Tev range that cannot be explained it terms of the simple HSZ SSC model (considering that second order IC scattering is negligible, see Massaro \cite{massarotesi}) because the IC peak falls for these sources around \(0.1\mbox{ GeV}\); this may be the case for BL Lacertae itself (Albert et al. \cite{albert2007b}) and possibly similar sources like M87 (Georganopoulos et al. \cite{gk2005}) and Cen A (Lenian et al. \cite{lenian}; Aharonian et al. \cite{aharonian2009}).

In all these cases the source may require more \emph{elaborate} structure, like decelerated relativistic outflows or sub-jet scenarios (see Georganopoulos \& Kazanas \cite{gk2003}; Tavecchio \& Ghisellini \cite{tavecchio2008}; Giannios et al. \cite{giannios}). We stress that the injection of a \emph{second}, nearly monoenergetic electron population is expected to be marked by a sudden \emph{increase} of the spectral curvature.

\section{Discussion and conclusions}

We propose that the sources of BL Lac type may be conveniently ordered in a \textit{succession} spanning from smooth variations of one or a few dominant SSC parameter, up to the appearance of truly different components that ultimately break through the limits of the simple HSZ SSC model. Even more so at increasing energies and frequencies, that would imply weaker and weaker \(\gamma\)-ray fluxes relative to X-ray, owing to KN cross section effects. This picture is supported by the following two lines of evidence.

First, a recent statistical study of Third EGRET Catalogue data (Mukherjee et al. \cite{mukherjee}; Casula \cite{victor}; Vagnetti, Casula \& Paggi in prep.) shows for BL Lac objects observed at \(30\mbox{ MeV}\div 30\mbox{ GeV}\) a weak \(\gamma\)-ray variability on average, compared to the FSRQ Blazars.
Also the time-structure functions for these two classes of objects indicate a similar trend, within the limitations of the sample.

Second, the first AGILE-GRID Catalogue of high confidence gamma-ray sources (Pittori et al. \cite{pittori}) apparently shows few if any new BL Lac sources other than those already detected by EGRET. This circumstance suggests the sources observed by AGILE to be more powerful than the average; it leads to expect for the majority of the sources either reduced flare activity in this band, or weak average fluxes. We relate these features to reduced Klein-Nishina cross section that tends to limit both the average fluxes and the flares.

With {\glast}, the first three months of observations (Abdo et al. \cite{abdo}) have yielded a substantial number of new BL Lac sources, and in particular a larger fraction of HBLs close to the ratio known from radio surveys (Nieppola et al. \cite{nieppola}). This is consistent with our expectations, considering the better instrumental sensitivity, especially at high energies generally favorable to the HBLs; for these sources we suggest to test the model's outcomes: \(C\) lower than \(S\), and generally small (though possibly fast) flux variations compared to the LBLs

We plan to check our theoretical predictions against the public availability of the first year data from {\glast} observations, and with simultaneous, multi-wavelength observations comparing the two basic subclasses of LBL and HBL of the BL Lac sources.

\begin{acknowledgements}
We thank our referee for useful comments and helpful suggestions. F. Massaro acknowledges the Foundation BLANCEFLOR Boncompagni-Ludovisi, n\'{e}e Bildt,
for the grant awarded him in 2009 to support his research.
\end{acknowledgements}

\begin{appendix}

\section{Log-parabolic spectra}\label{calcoli}

In this Appendix we show how electron populations with a log-parabolic energy distribution of the form expressed by Eq. (\ref{logpar1}), that is,
\begin{equation}\label{logpar}
N(\gamma)=N_0\,{\left({\frac{\gamma}{\gamma_0}}\right)}^{-s-r\log{\left({\frac{\gamma}{\gamma_0}}\right)}},
\end{equation}
emit log-parabolic spectra via the SSC process. The related particle synchrotron emissivity
\begin{equation}
j_\nu^s = \int{d\gamma\,N(\gamma)\,\frac{dP_s}{d\nu}}
\end{equation}
is easily computed on using the close approximation to the single particle emission in the shape of a delta-function (see. Rybicki \& Lightmann \cite{rl}), that is, \(\frac{dP_s}{d\nu}\approx P_s\,\delta\,{\left( {\nu-\gamma^2\,\nu_c}\right) }\) with \(P_s={1}/{6\pi}\,\sigma_T\,\gamma^2\,c\,{B^2}\) and \(\nu_c\approx 1.22\, {10}^6\, B\mbox{ Hz}\) (with \(B\) measured in Gauss) is the synchrotron critical frequency. This leads to (Massaro et al. \cite{massaro2004a}) a log-parabolic differential flux
\begin{equation}
F_\nu^s \approx F_0{\left( {\frac{\nu}{\nu_0}}\right)}^{-a_s-b_s\log{\left( {\frac{\nu}{\nu_0}}\right) }},
\end{equation}
and to a SED again of log-parabolic shape
\begin{equation}
S_\nu^s =\nu F_\nu^s\approx S_0^s\,{\left( {\frac{\nu}{\nu_0}}\right)}^{-(a_s-1)-b_s\log{\left( {\frac{\nu}{\nu_0}}\right) }};
\end{equation}
its slope at the synchrotron reference frequency \(\nu_0\) is given in terms of \(s\), by
\begin{equation}
a_s\approx\frac{s-1}{2},
\end{equation}
the spectral curvature by
\begin{equation}\label{curvaturasincro}
b_s\approx\frac{r}{4},
\end{equation}
and the peak value \(S'\propto R^3\,B^2\,n\,\gamma_p^2\,\sqrt{r}\) occurs at a frequency \(\xi'\propto B\,\gamma_p^2\,{10}^{\frac{1}{r}}\) (\(\gamma_p\), \(n\), \(B\) and \(R\) are defined in \S \ref{spectra} of the main text).

For IC radiation in the Thomson regime we may write to a fair approximation \(\frac{dP_c}{d\nu}\approx P_c\,\delta\,{\left( {\nu-\frac{4}{3}\,\gamma^2\,\xi'}\right) }\) (see Rybicky \& Lightmann \cite{rl}) where \(P_c=\frac{4}{3}\,\sigma_T\,\gamma^2\,c\,\epsilon_\nu\) is the  power radiated by a single-particle IC scattering in the Thomson regime, having denoted with \(\epsilon_\nu\propto R\,B^2\,n\,\gamma_p^2\) the synchrotron radiation density\footnote{Note that in Thomson regime no direct relation appears between electrons emitting peak synchrotron photons and those responsible of IC peak radiation (see Tavecchio et al. \cite{tavecchio1998}).}
. We obtain once again a log-parabolic SED
\begin{equation}
S_\nu^c \approx S_0^c\,{\left( {\frac{\nu}{\hat{\nu}_0}}\right)}^{-(a_c-1)-b_c\log{\left( {{\nu}/{\hat{\nu}_0}}\right) }},
\end{equation}
where the slope at the IC reference frequency \(\hat{\nu}_0\) is given by
\begin{equation}
a_c\approx\frac{s-1}{2},
\end{equation}
and the spectral curvature reads
\begin{equation}\label{curvaturathomson}
b_c\approx\frac{r}{4}.
\end{equation}
The peak value \(C'\propto R^4\,B^2\,n^2\,\gamma_p^4\,\sqrt{r}\) is attained at a frequency \(\epsilon'\propto B\,\gamma_p^4\).

For the Klein-Nishina (KN) regime instead it necessary to consider the convolution
\begin{equation}
 j_\nu^c =h
\int{
d\gamma\,N(\gamma)
\int{d\tilde\nu\,\nu\,N_{\tilde\nu}\,K(\nu,\tilde\nu,\gamma)}
}
\end{equation}
where \(\tilde\nu\) and \(\nu\) are the electron frequencies before and after the scattering, respectively, \(N_{\tilde\nu}\) is the number spectrum of seed photons, and \(K(\nu,\tilde\nu,\gamma)\) is the full Compton kernel (Jones \cite{jones}). Only in the extreme KN regime one may again approximate
\begin{equation}
K(\nu,\tilde\nu,\gamma)\approx\frac{1}{\gamma^2}\,\delta\,{\left( {1-\frac{h\nu}{\gamma m c^2}}\right) }\,;
\end{equation}
on approximating \(N_{\tilde\nu}\) with its mean value for a homogeneous, spherical optically thin source
\begin{equation}
 N_{\tilde\nu}\approx \left\langle {N_{\tilde\nu}}\right\rangle =\frac{3}{4}\frac{R}{c}\frac{j_{\tilde\nu}^s}{h\tilde\nu},
\end{equation}
one obtains again a SED with a log-parabolic shape
\begin{equation}
S_\nu^c \approx S_0^c\,{\left( {\frac{\nu}{\hat{\nu}_0}}\right)}^{-(a_c-1) -b_c\log{\left( {{\nu}/{\hat{\nu}_0}}\right) }}\,.
\end{equation}
Now the slope at \(\hat{\nu}_0\)
\begin{equation}
a_c\approx s,
\end{equation}
is \emph{steeper}, and the spectral curvature
\begin{equation}\label{curvaturakn}
b_c\approx r
\end{equation}
 is \emph{larger} than in the Thomson regime. The peak value \(C'\propto R^4\,B\,n^2\,\sqrt{r}\) occurs at a frequency \(\epsilon'\propto \gamma_p\,{10}^{-\frac{1}{2r}}\).

The transition between the two regimes occurs when \(2\,\gamma_{max}\, h\,\xi'\approx m c^2\), where \(\gamma_{max}=\sqrt{\xi'/\nu_c}\), that is, when
\begin{equation}
\xi_T\approx 7.15\,{10}^{15}\,{\left({\frac{B}{0.1\mbox{ G}}}\right)}^{\frac{1}{3}}\,
{\left({\frac{\delta}{10}}\right)}\,(1+z)^{-1}\mbox{ Hz}\,
\end{equation}
holds, or equivalently
\begin{equation}
\xi_T\approx 1.96\,{10}^{16}\,{\left({\frac{\gamma_p}{{10}^4}}\right)}^{-1}\,
{10}^{\frac{1}{2}\left({1-\frac{1}{5b}}\right)}\,
{\left({\frac{\delta}{10}}\right)}\,(1+z)^{-1}\mbox{ Hz}\, .
\end{equation}

We close these calculations with two remarks. First, the (primed) quantities used here refer to the rest frame of the emitting region, while the observed (unprimed) quantities must be multiplied by powers of the beaming factor
\begin{equation}
 \delta=\frac{1}{\Gamma \left( {1-\beta\,\cos{\theta}}\right)},
\end{equation}
\(\Gamma\) being the bulk Lorentz factor of the relativistic electron flow in the emitting region, \(\theta\) the angle between its velocity \(\varv\) and the line of sight, and \(\beta=\varv/c\); so we obtain \(S=S'\,\delta^4\), \(C=C'\,\delta^4\), and \(\xi=\xi'\,\delta\), \(\epsilon=\epsilon'\,\delta\).
Second, we note that the actual specific powers \(\frac{dP}{d\nu}\) radiated by a single particle differ from delta-function shape, slightly for the synchrotron emission and considerably for the IC radiation in the Thomson regime (see discussion in Rybicky \& Lightmann \cite{rl}). However, the above spectral shapes still approximatively apply, as confirmed by numerical simulations (Massaro \cite{massarotesi}); in detail, the convolution with a broader single particle power yields a \emph{less} curved spectrum, so that equations \ref{curvaturasincro} and \ref{curvaturathomson} become \(b_s \approx r/5\) (Massaro et al. \cite{massaro2006}) and \(b_c\approx r/10\), respectively.

\section{Particle acceleration processes}\label{accelerazione}

Here we derive a log-parabolic electron energy distribution \(N\left({\gamma ,t}\right)\) from a kinetic continuity equation of the Fokker-Planck type; following Kardashev (\cite{kardashev}), in the jet rest frame this reads
\begin{equation}\label{fp}
 \frac{\partial N}{\partial t}=-
\lambda_1 (t)\, \frac{\partial}{\partial\gamma}\left({\gamma\,N}\right)+
\lambda_2 (t)\, \frac{\partial}{\partial\gamma}\left({\gamma^2\,\frac{\partial N}{\partial\gamma}}\right),
\end{equation}
where \(\gamma\, m\, c^2\) is the particle energy, \(t\) denotes time, and \(\lambda_1\) and \(\lambda_2\) describe systematic and stochastic acceleration rates, occurring on timescales \(t_1=1/\lambda_1\) and \(t_2=1/\lambda_2\), respectively. For example, in the picture of Fermi accelerations (e.g., Vietri \cite{vietri}), the accelerations rates \(\lambda_1\) and \(\lambda_2\) can be expressed in terms of physical quantities related to processes occurring in shocks; in this framework a plane shock front of thickness \(\ell_s\) moves with speed \(V_s\) and gas clouds of average size \(\ell\) move downstream of the shock with speed \(V\); it is found that \(\lambda_1=V_s/\ell_s\) and \(\lambda_2=V^2/2c\ell\) hold (see Kaplan \cite{kaplan}; Paggi \cite{paggi}), \(\lambda_2\ll\lambda_1\). On the other hand, numerical values of \(\lambda_1\) and \(\lambda_2\) are directly derived from the emitted spectrum as shown later.

The Fokker-Planck equation (\ref{fp}) describes the \emph{evolution} of the electron distribution function; with an initially mono-energetic distribution in the form of a delta-function \(N(\gamma ,0) = n \,\delta(\gamma -\gamma_0)\) (\(n\) is the initial particle number density) the solution at subsequent times \(t\) takes the form of the log-parabolic energy distribution assumed in Eq. (\ref{logpar1}) of the main text (see also \ref{logpar}), reading
\begin{equation}\label{logpar2}
 N\left({\gamma ,t}\right)= N_0\, {\left({\frac{\gamma}{\gamma_0}}\right)}^{-s-r\log{\left({\frac{\gamma}{\gamma_0}}\right)}}.
\end{equation}
Here the time depending slope at \(\gamma=\gamma_0\) is given by
\begin{equation}
s=\frac{1}{2}\left( {1-\frac{\int{dt\,\lambda_1}}{\int{dt\,\lambda_2}}}\right)\, ;
\end{equation}
meanwhile, the curvature of \(N(\gamma)\) driven by the diffusive (stochastic) term in Eq. (\ref{fp}), \textit{irreversibly} decreases in time after
\begin{equation}
r=\frac{\ln{10}}{4\int{dt\,\lambda_2}}
\end{equation}
from the large initial values corresponding to the initially mono-energetic distribution. Correspondingly, the time-dependent height at \(\gamma=\gamma_0\) follows
\begin{equation}
N_0=\frac{1}{2\sqrt{\pi}}\frac{n}{\gamma_0}\frac{1} {\sqrt{\int{dt\,\lambda_2}}}\, {\exp{\left[{-\frac{{\left({\int{dt\,\lambda_1}+\int{dt\,\lambda_2}}\right)}^2}
{4\int{dt\,\lambda_2}}}\right]}}.
\end{equation}

To wit, Eq. (\ref{logpar2}) describes the evolution of the electron distribution, growing broader and \emph{broader} under the effect of stochastic acceleration, while its peak \emph{moves} from \(\gamma_0\) to the current position
\begin{equation}
\gamma_{M}=\gamma_0\, {e}^{\int{dt\,(\lambda_1-\lambda_2)}}
\end{equation}
under the contrasting actions of the systematic and stochastic accelerations. An important quantity to focus on for the emission properties is the r.m.s. energy
\begin{equation}
 \gamma_p\equiv\sqrt{\frac{\int{\gamma^2\, N(\gamma)\, d\gamma}}{\int{ N(\gamma)\, d\gamma}}}=\gamma_0\, {e}^{\int{(\lambda_1+3\lambda_2)\,dt}}=\gamma_0\, {10}^{\frac{2-s}{2r}}=\gamma_{M}\,{e}^{{4\int{\lambda_2\,dt}}},
\end{equation}
which is also the position for the peak of the distribution \(\gamma^2\, N(\gamma)\).

We have already derived in Appendix \ref{calcoli} the shapes of the spectra (synchrotron, and IC in both the Thomson and KN regimes) emitted by the distribution given in Eq. (\ref{logpar1}); here we stress the time dependence of their main spectral features. We can write for the synchrotron emission\footnote{Note that singular behaviors of \(S\) and \(C\) for \(t=0\) are a consequence of the (differential) definition of the SED as \(\nu F_\nu\), and relate to the initial singularity of the particle energy distribution. Integrated quantities like \(F=\int{d\nu\,F_\nu}\) behave regularly.}
\begin{equation}\label{fluxsincro}
S\propto \frac{{e}^{2\int{\left({\lambda_1+3\lambda_2}\right)dt}}}{\sqrt{\int{\lambda_2\,dt}}}\propto \gamma_p^2\sqrt{r},\qquad
\xi\propto {e}^{2\int{\left({\lambda_1+5\lambda_2}\right)dt}}\propto\gamma_p^2\,{10}^{\frac{1}{r}}\, ;
\end{equation}
for IC emission we have in the Thomson regime
\begin{equation}
C\propto \frac{{e}^{4\int{\left({\lambda_1+3\lambda_2}\right)dt}}}{\sqrt{\int{\lambda_2\,dt}}}\propto\gamma_p^4\sqrt{r},\qquad
\epsilon\propto {e}^{4\int{\left({\lambda_1+5\lambda_2}\right)dt}}\propto\gamma_p^4\, ,
\end{equation}
and in the extreme KN regime
\begin{equation}\label{fluxICKN}
C\propto  \frac{1}{\sqrt{\int{\lambda_2\,dt}}}\propto \sqrt{r},\qquad
\epsilon\propto {e}^{\int{\left({\lambda_1+\lambda_2}\right)dt}}\propto\gamma_p\,{10}^{-\frac{1}{2r}}.
\end{equation}
\begin{figure}[ht]
 \centering
 \includegraphics[scale=0.45]{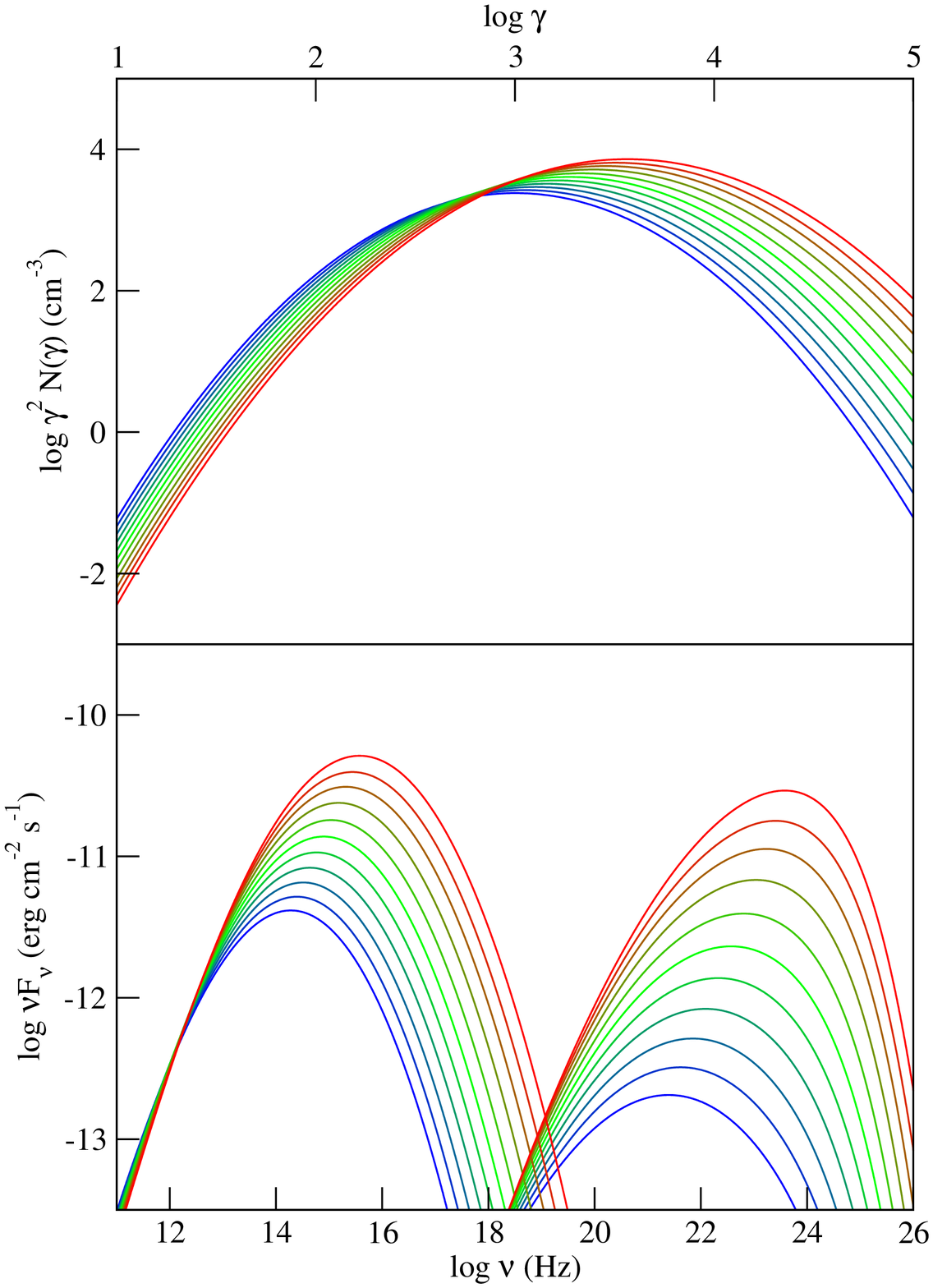}
\caption{Example of time evolution of the \(\gamma^2\,N(\gamma)\) distribution (with peak energy \(\gamma_p\)) due to stochastic and systematic accelerations (upper panel) with \(\lambda_1 = 5\, {10}^{-3}\mbox{ days}^{-1}\), \(\lambda_2 = 5\, {10}^{-4}\mbox{ days}^{-1}\) from the initial value \(\gamma_p={10}^3\), and of the related evolution for the SSC SEDs (lower panel). In terms of the stochastic acceleration time \(\tau_2\), the time interval between each pair of lines is \(t_2-t_1={10}^{-2} \tau_2\), corresponding to an observed time interval of about two days.}
\label{evolthm}
 \end{figure}
\begin{figure}[ht]
 \centering
 \includegraphics[scale=0.45]{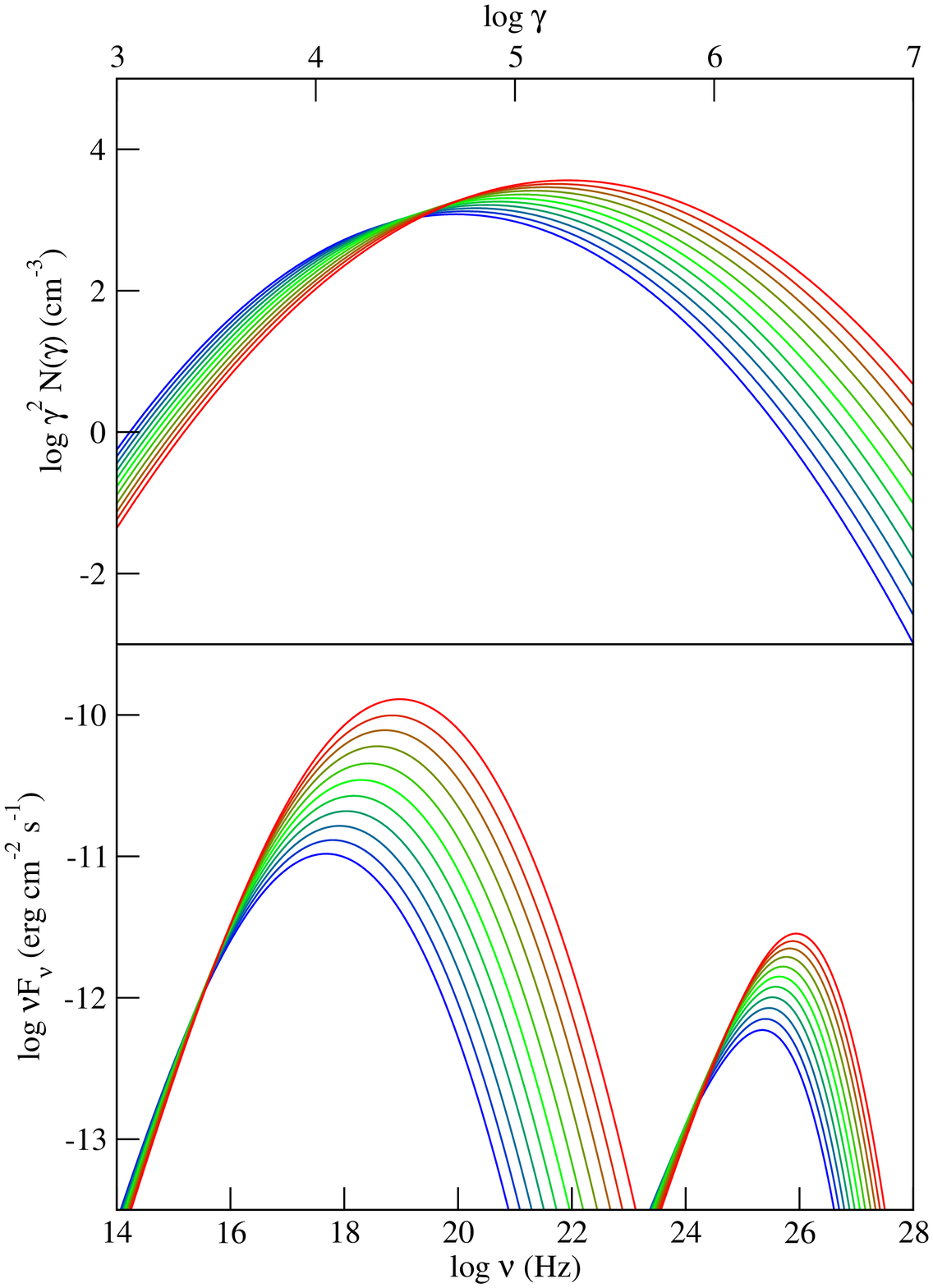}
\caption{Example of time evolution of the \(\gamma^2\,N(\gamma)\) distribution (with peak energy \(\gamma_p\)) due to stochastic and systematic accelerations (upper panel) with \(\lambda_1 = 5\, {10}^{-3}\mbox{ days}^{-1}\), \(\lambda_2 = 5\, {10}^{-4}\mbox{ days}^{-1}\) from the initial value \(\gamma_p=5\,{10}^4\), and of the related evolution for the SSC SEDs (lower panel). In terms of the stochastic acceleration time \(\tau_2\), the time interval between each pair of lines is \(t_2-t_1={10}^{-2} \tau_2\), corresponding to an observed time interval of about two days.}
\label{evolkn}
 \end{figure}
Note that during flares, since \(\lambda_1\gg \lambda_2\), we expect the curvature to vary so little that we can therefore approximatively write \(S\propto\xi\) (see main text).

Now we focus the above relations for the simple case of time independent \(\lambda_1\) and \(\lambda_2\), when \(\int{\lambda_{1,2}\, dt}\approx\lambda_{1,2}\, t\); then we have
\begin{equation}\label{curvvstime}
s=\frac{1}{2}\left( {1-\frac{\lambda_1}{\lambda_2}}\right),\qquad
r=\frac{\ln{10}}{4\,\lambda_2\,t}\approx  \frac{0.58}{\lambda_2\,t},
\end{equation}
while for the r.m.s. energy we have
\begin{equation}\label{r.m.s.}
\gamma_p=\gamma_0 \,{e}^{\left({\lambda_1+3\lambda_2}\right)t};
\end{equation}
so for the peak frequency and the spectral curvature we have
\begin{equation}\label{curvvsxi}
\log{\left({\frac{\xi}{\xi_0}}\right)}={2 \left({\lambda_1+5\lambda_2}\right)t}, \qquad b_s = \frac{1}{10}\left({5+\frac{\lambda_1}{\lambda_2}}\right)
\frac{1}{\log{\left({\frac{\xi}{\xi_0}}\right)}},
\end{equation}
where \(\xi_0\) is a normalization frequency. It is seen that soon after the injection the curvature \(b_s\) (proportional to \(r\)) \emph{drops} rapidly, then progressively \emph{decreases} more and more gently, while the peak frequency still \emph{increases}.

The value of \(\lambda_2\) can be evaluated from observing the synchrotron spectral curvatures \(b_2\) and \(b_1\) at two times \(t_2\) and \(t_1\), respectively (recall that \(b_s\approx r/5\)); denoting with \( t_2 - t_1\) this time interval, we have
\begin{equation}
\lambda_2=\frac{0.58}{{t_2 - t_1}}\left( {\frac{1}{r_2}-\frac{1}{r_1}}\right) \frac{1+z}{\delta};
\end{equation}
on the other hand, form observing the related synchrotron peaks \(\xi_2\) and \(\xi_1\), the value of \(\lambda_1\) can be evaluated as
\begin{equation}
\lambda_1=\left[{\frac{1}{2\left({t_2 - t_1}\right)}\ln{\left( {\frac{\xi_2}{\xi_1}}\right)}-\frac{2.88}{{t_2 - t_1}}
\left( {\frac{1}{r_2}-\frac{1}{r_1}}\right)}\right] \frac{1+z}{\delta}\, .
\end{equation}
For example, in the case of {\mrkb} in the states of 7 and 16 April 1997, we obtain (on assuming \(B\approx\mbox{const}\)) \(\lambda_1 = (2.3\pm 1.1)\mbox{ yr}^{-1}\) and \(\lambda_2 = (1.8\pm 1.7){10}^{-1}\mbox{ yr}^{-1}\), corresponding to acceleration times \(\tau_1=1/\lambda_1=(4.3\pm 2.0){10}^{-1}\mbox{ yr}\) and \(\tau_2=1/\lambda_2=(5.5\pm 5.0)\mbox{ yr}\). Note that with the current data the evaluations of \(\lambda_1\) and \(\lambda_2\) turn out to be affected by uncertainties considerably larger than the single curvatures \(b_1=0.161\pm 0.007\) and \(b_2=0.148\pm 0.005\).

If the total energy available to the jet is limited (e.g., by the BZ limit, see text) we expect that \(\int{dt\,\lambda_1}\) and \(\int{dt\,\lambda_2}\) cannot grow indefinitely, but are to attain a \emph{limiting} value. At low energies where \(\int{dt\,\lambda_1}\gg\int{dt\,\lambda_2}\) holds, we have \(S\propto \xi\) and \(C\propto \epsilon\) as before; at higher energies when \(\int{dt\,\lambda_1}\) reaches its limit, \(\int{dt\,\lambda_1}\ll\int{dt\,\lambda_2}\) holds, leading to \(S\propto \xi^{0.6}\) and \(C\propto \epsilon^{0.6}\). Eventually also \(\int{dt\,\lambda_2}\) reaches its limit, and both the fluxes and the peak frequencies cannot grow any more.

We add that the effect of radiative cooling is to reshape the distribution by moving high energy particles to lower energies; a corresponding deformation takes place in the emitted spectra at their high energy end, leaving nearly unaffected the peak fluxes for long cooling times. Then a term \(\eta\, \frac{\partial}{\partial\gamma}\left({\gamma^2\,N}\right)\) is added on the r. h. s. of Eq. (\ref{fp}), where for synchrotron radiative cooling \(\eta=\frac{1}{6\pi}\,\sigma_T\,c\,\beta^2\,{B^2}\) holds. The equation can still be solved on using the Kaplan approximation (Kaplan \cite{kaplan}) when \(\lambda_1\) and \(\lambda_2\) are time independent; the solution is a power series in the parameter \(q={\eta\,\gamma_0}/\left({{2\,\lambda_2}}\right)\ll 1\), that reads
\begin{equation}
 N\left({\gamma ,t}\right)\approx N_0\, {\left({\frac{\gamma}{\gamma_0}}\right)}^ {-s-r\log{\left({\frac{\gamma}{\gamma_0}}\right)}}\,
\left[{1+q\,\left({1-\frac{\gamma}{\left\langle {\gamma}\right\rangle} }\right)}\right]\, ,
\end{equation}
where
\begin{equation}
\left\langle {\gamma}\right\rangle=\gamma_0\, {10}^{\frac{3-2s}{4r}}
\end{equation}
is particle mean energy.

\section{Prefactors for Eqs. (\ref{prima}) - (\ref{ultima})}\label{constraints}

The HSZ SSC model is, as stated before, characterized by five parameters: the r.m.s. particle energy \(\gamma_p\), the particle density \(n\), the magnetic field \(B\), the size of the emitting region \(R\) and the beaming factor \(\delta\). So the model may be constrained by five observables that we denote with
\begin{equation}
S_i\equiv\frac{S}{{10}^{i}\frac{\mbox{ erg}}{\mbox{ cm}^2\mbox{ s}}}\, ,
\qquad
C_j\equiv\frac{C}{{10}^{j}\frac{\mbox{ erg}}{\mbox{ cm}^2\mbox{ s}}}\, ,
\end{equation}
\begin{equation}
\xi_k\equiv\frac{\xi}{{10}^{k}\mbox{ Hz}}\, ,
\qquad
\epsilon_h\equiv\frac{\epsilon}{{10}^{h}\mbox{ Hz}}\, ,
\end{equation}
where the indexes \(i,j,k,h\) express the normalizations as demonstrated below; in addition, we denote with \(\Delta t_d\) the time in days for the source variations,  and with \(D\) the distance of the source in Gpc.

In the Thomson regime we find
\begin{eqnarray}
B &=&{10}^{-1}\times\\
&\times&\left[{{b}^{\frac{1}{8}}\,{D}^{-\frac{1}{2}}\,{\left({1+z}\right)}^{\frac{1}{2}}\,
{\xi_{14}}^{3}{C_{-11}}^{\frac{1}{4}}\,{\Delta t_{d}}^{\frac{1}{2}}\,
{\epsilon_{22}}^{-\frac{3}{2}}\,{S_{-11}}^{-\frac{1}{2}}}\right]\mbox{ G} \nonumber
\end{eqnarray}
\begin{eqnarray}
\delta&=&13.5\times \\
&\times&\left[{{D}^{\frac{1}{2}}\,{b}^{-\frac{1}{8}}\,{\left({1+z}\right)}^{\frac{1}{2}} \,
{\epsilon_{22}}^{\frac{1}{2}}\,{S_{-11}}^{\frac{1}{2}}\,
{\xi_{14}}^{-1}\,{C_{-11}}^{-\frac{1}{4}}{\Delta t_{d}}^{-\frac{1}{2}}}\right]\nonumber
\end{eqnarray}
\begin{eqnarray}
R&=&3.5\,{10}^{16}\times  \\
&\times&\left[{{D}^{\frac{1}{2}}\, {b}^{-\frac{1}{8}}\,{\left({1+z}\right)}^{-\frac{1}{2}}\,
{\epsilon_{22}}^{\frac{1}{2}}\,{S_{-11}}^{\frac{1}{2}}\,{\Delta t_{d}}^{\frac{1}{2}}\,
{\xi_{14}}^{-1}\,{C_{-11}}^{-\frac{1}{4}}}\right]\mbox{ cm}\nonumber
\end{eqnarray}
\begin{eqnarray}
n&=&5.7\times  \\
&\times& \left[{{b}^{\frac{1}{8}}\,{D}^{\frac{1}{2}}\,{\left({1+z}\right)}^{\frac{1}{2}}\,{10}^{\left({\frac{1}{5b}-1}\right)}\,
{\xi_{14}}^{2}\,{C_{-11}}^{\frac{5}{4}}\,
{\epsilon_{22}}^{-\frac{3}{2}}\,{S_{-11}}^{-\frac{3}{2}}\,{\Delta t_{d}}^{-\frac{1}{2}}}\right]\mbox{ cm}^{-3}\nonumber
\end{eqnarray}
\begin{equation}
\gamma_p=2.7\,{10}^3\,\left[{ {10}^{\frac{1}{2}\left({1-\frac{1}{5b}}\right)}\,
{\epsilon_{22}}^{\frac{1}{2}}\,{\xi_{14}}^{-\frac{1}{2}}}\right]\,.
\end{equation}

In the extreme KN regime we obtain
\begin{eqnarray}
B&=&1.7\,{10}^{-2}\times  \\
&\times&\left[{{D}^{\frac{2}{5}}\,{b}^{\frac{1}{10}}\,{(1+z)}^{-\frac{2}{5}}\,
{\xi_{18}}^{\frac{2}{5}}\,{S_{-10}}^{\frac{2}{5}}\,
{\epsilon_{26}}^{-\frac{8}{5}}\,{C_{-11}}^{-\frac{1}{5}}\,{\Delta t_d}^{-\frac{2}{5}}}\right]\mbox{ G}\nonumber
\end{eqnarray}
\begin{eqnarray}
\delta&=&4.2\times  \\
&\times&\left[{{D}^{\frac{2}{5}}\,{b}^{-\frac{1}{10}}\,
{10}^{\frac{4}{5}\left({\frac{1}{5b}-1}\right)}\,
{(1+z)}^{\frac{3}{5}}\,
{\epsilon_{26}}^{\frac{2}{5}}\,{S_{-10}}^{\frac{2}{5}}\,
{\xi_{18}}^{-\frac{3}{5}}\,{C_{-11}}^{-\frac{1}{5}}\,{\Delta t_d}^{-\frac{2}{5}}}\right]\nonumber
\end{eqnarray}
\begin{eqnarray}
R&=&1.1\, {10}^{16}\times  \\
&\times&\left[{{D}^{\frac{2}{5}}\,{b}^{-\frac{1}{10}}\,{10}^{\frac{4}{5}\left({\frac{1}{5b}-1}\right)}\,{(1+z)}^{-\frac{2}{5}}\,
{\epsilon_{26}}^{\frac{2}{5}}\,{S_{-10}}^{\frac{2}{5}}\,{\Delta t_d}^{\frac{3}{5}}\,
{\xi_{18}}^{-\frac{3}{5}}\,{C_{-11}}^{-\frac{1}{5}}}\right]\mbox{ cm}\nonumber
\end{eqnarray}
\begin{eqnarray}
n&=&9.5\,{10}^2\times  \\
&\times&\left[{{b}^{\frac{1}{5}}\,{D}^{-\frac{4}{5}}\,{10}^{\frac{13}{5}\left({1-\frac{1}{5b}}\right)}\,{(1+z)}^{\frac{4}{5}}\,
{\xi_{18}}^{\frac{11}{5}}\,{C_{-11}}^{\frac{7}{5}}\,
{\epsilon_{26}}^{-\frac{4}{5}}\,{S_{-10}}^{-\frac{9}{5}}\,{\Delta t_d}^{-\frac{1}{5}}}\right]\mbox{ cm}^{-3}\nonumber
\end{eqnarray}
\begin{eqnarray}
\gamma_p&=&6.1\,{10}^{5}\times  \\
&\times&\left[{{b}^{\frac{1}{10}}\,{D}^{-\frac{2}{5}}\,{10}^{\frac{3}{10}\left({1-\frac{1}{5b}}\right)}\,{(1+z)}^{\frac{2}{5}}\,
{\xi_{18}}^{\frac{3}{5}}\,{\epsilon_{26}}^{\frac{3}{5}}\,{C_{-11}}^{\frac{1}{5}}\,{\Delta t_d}^{\frac{2}{5}}\,
{S_{-10}}^{-\frac{2}{5}}}\right]\,.\nonumber
\end{eqnarray}

\section{KN correlations in the decay stage}\label{newappendix}

Here we briefly comment on the decay phase of a flare, when fluxes and frequencies decrease; the aim is at understanding the quadratic correlations between the \(\gamma\)-ray and the X-ray fluxes observed in HBL objects like Mrk421 and Mrk501 (see Katarzynski et al. \cite{katarzynsky}; Fossati et al. \cite{fossati}). Electron energy losses by adiabatic expansion (at a constant rate) can be accounted for by adding to the Fokker-Planck equation a negative term that combines with the systematic acceleration rate (Kardashev \cite{kardashev}) into an overall coefficient \(\lambda_1 \rightarrow \lambda_1 -\frac{1}{t}\).

Correspondingly, the size of the source grows as \(R\propto t\);  with the total particle number remaining constant, in the extreme Klein-Nishina regime Eqs.  \ref{fluxsincro} and \ref{fluxICKN} may be rewritten as
\begin{equation}
S\propto \exp{\left[{2\int{\left({\lambda_1-\frac{1}{t}}\right)dt}}\right]},
\end{equation}
\begin{equation}
C\propto \exp{\left[{-2\int{\frac{1}{t}\,dt}}\right]},
\end{equation}
on neglecting the stochastic acceleration contribution.

If both front shock thickness and gas clouds size scale as \(R(t)\) then \(\lambda_1=1/(A\, t)\) holds. With \(A\geq 1\) both the synchrotron and the IC flux will decrease. The particular choice \(A=2\) yields a quadratic correlation between the IC and the synchrotron peaks, that is, \(\Delta \log{C}=2\,\Delta \log{S}\). One should not quail at the fine tuning corresponding to \(A=2\) literally, because lower values of \(1<A<2\) are more effective in a mild KN regime. On the other hand, A cannot be large lest the shock is stalled.

\end{appendix}

\end{document}